\documentclass[useAMS,usenatbib]{mn2e}
\usepackage{graphicx}

\title[An original constraint on the Hubble constant : h$>$0.74]{An original constraint on the Hubble constant : h$>$0.74}
\author[A. Barrau, A. Gorecki and J. Grain]{A. Barrau$^{1}$\thanks{E-mail:
Aurelien.Barrau@cern.ch}, A. Gorecki$^{1}$ and J. Grain$^{2}$\\
$^{1}$Laboratoire de Physique Subatomique et de Cosmologie,
 Universit\'e Joseph Fourier, CNRS/IN2P3, INPG\\
53, avenue des Martyrs, 38026 Grenoble cedex, France\\
$^{2}$Laboratoire AstroParticule et Cosmologie, Universit\'e Paris 7, CNRS/IN2P3 \\
10, rue Alice Domon et L\'eonie Duquet, 75205 Paris cedex 13, France}
\begin{document}

\date{Accepted xxxx Received xxxx; in original form xxxx}

\pagerange{\pageref{firstpage}--\pageref{lastpage}} \pubyear{2008}

\maketitle

\label{firstpage}

\begin{abstract}
The Hubble parameter $H_0$ is still not very well measured. Although the Hubble
Key Project, Chandra and WMAP gave good estimates, the uncertainties remain
quite large. In this brief report, we suggest an original and independent method to derive a
lower limit on $H_0$ using the absorption of very high energy gamma-rays by
the cosmic infrared background. With conservative a hypothesis, we obtain
$H_0>74$~km.s$^{-1}$.Mpc$^{-1}$ at the 68\%
confidence level, which favors the upper end of the
intervals allowed by dedicated experiments. 
\end{abstract}


\section*{Introduction}

Gamma-rays in the $10^{11}-10^{13}$~eV range are observed from distant Active Galactic 
Nuclei (AGN). Those
very high-energy photons will be absorbed by the Cosmic Infrared Background 
(CIB) due to electron-positron pair production. In the higher energy range
(typically around 10-20~TeV), the absorption is expected to be so high that the
difference between the source spectrum and the measured spectrum becomes very
large. As the CIB spectral distribution is now quite correctly known (at least with
lower limits) it is possible to compute the intrinsic spectrum of a given
AGN as a function of the integrated CIB density crossed along the line-of-sight
through an unfolding procedure. 
The absorption being proportional to $e^{1/H_0}$, it is then possible to derive a sensitive 
lower limit on the Hubble parameter so as to exclude unfolded spectra which would be 
explicitly unphysical. The idea is very simple~: the smaller the Hubble constant, the 
larger the distance to the source, the larger the number of CIB photons
crossed, and the higher the unfolding coefficients. As the unfolded ({\it
i.e.} intrinsic) spectrum shape will inevitably become unacceptable for low
enough
Hubble parameters, this approach allows to derive a lower limit on $H_0$.
The obtained bound is relevant when compared with current estimates, although the main
aim of this brief article is to open this field and outline the scheme of the method.
In the
first section, the latest measurements of the CIB density are reviewed,
including some recently revised analysis. In the second section, the gamma-ray data
relevant for this study are presented together with the associated uncertainties.
In the third section, the physical processes involved in the gamma-ray absorption
mechanism are explained and the basic features of the proposed method are
underlined. The fourth section deals with the details of the unfolding procedure
and with the physical criteria used to reject "unphysical" unfolded spectra. In 
the fifth section, the Monte-Carlo method used to derive statistically meaningful
bounds is described and the results are
given. Finally, several developments that could be 
expected in the future are  outlined in the last section.\\

\section{The cosmic infrared background density}

Our understanding of the early epochs of galaxies has recently
increased thanks to the observational evidences provided by 
UV/Visible/Near-IR, far-IR and submillimeter surveys of high-redshift objects. 
In a consistent scenario, galaxy formation and evolution
can also be constrained by the background radiation which 
is produced by the line-of-sight accumulation of all extragalactic sources. The Cosmic Infrared
Background (CIB) is basically the relic emission at wavelengths between a few microns and
millimeters of the formation and evolution of galaxies of all type and star-forming systems
(see, {\it e.g.}, Puget {\it et al.} \cite{puget}, Hauser {\it et al.} \cite{hauser}, Lagache 
{\it et al.} \cite{lagache}, Gispert {\it et al.} \cite{gispert}, Hauser \& Dwek
\cite{hauserdwek} and Kashlinsky \cite{kashlinsky}). The near-IR CIB arises mainly from the
stellar component of galaxies and probes their evolution at early
times. The mid- and far-IR CIB originates from dusty galaxies reprocessing stellar light and
other energetic output.\\

Observationally, the CIB is difficult to distinguish from the generally brighter foregrounds
contributed by the local matter within the solar system, the stars and the interstellar medium
of the Galaxy. However, the situation has dramatically improved in the last
decade. In this study, we have used the most accurate and "up-to-date" estimates of the 
CIB density, as described, {\it e.g.}, in Dole {\it et al.} \cite{dole06} with linear
interpolations (in log-log scale) between the measurements. The doubtful measurements have deliberately been
ignored. The number of points is now large enough to make this
approach quite accurate. The measurements used in our analysis are the following:
\begin{itemize}
\item 0.3 $\mu$m : 10$\pm 5$ nW.m$^{-2}$.sr$^{-1}$, from Bernstein {\it et al.}
\cite{bernstein} corrected by Matilla \cite{matilla}. The estimate is based on the Hubble Space
Telescope (HST) broadband CCD photometry and the 2.5 m du Pont telescope from Lac Campanas
Observatory (LCO).
\item 1.25 $\mu$m : 28$\pm 15$ nW.m$^{-2}$.sr$^{-1}$, from Wright \cite{wright} and Cambresy
{\it et al.} \cite{cambresy}. The estimate relies on the Diffuse Infrared Background Experiment
(DIRBE) and 2 Micron All Sky Survey (2MASS) data.
\item 2.2 $\mu$m : 22.4$\pm 6$ nW.m$^{-2}$.sr$^{-1}$,  from Gorjian {\it et al.}
\cite{gorjian} using DIRBE and {\it Lick} data.
\item 3.3 $\mu$m : 11.1$\pm 3.3$ nW.m$^{-2}$.sr$^{-1}$, from the same analysis.
\item 15 $\mu$m : 3.3$\pm 1.3$ nW.m$^{-2}$.sr$^{-1}$, from Elbaz {\it et al.} \cite{elbaz}. This lower limit
is based on galaxy counts with the Infrared Space Observatory Camera (ISOCAM).
\item 24 $\mu$m : 2.7$\pm 1.1$ nW.m$^{-2}$.sr$^{-1}$, from Papovich {\it et al.} \cite{papovich}. The
limit comes from galaxy counts with the Multiband Imaging Photometer for {\it Spitzer} 
(MIPS). 
\item 70 $\mu$m : 7.1$\pm 1.0$ nW.m$^{-2}$.sr$^{-1}$, from Dole {\it et al.} \cite{dole06}, 
based on MIPS data with more than 19000 24 $\mu$m sources stacked with $S_{24}>60~\mu$ Jy.
\item 160 $\mu$m : 13.4$\pm 1.7$ nW.m$^{-2}$.sr$^{-1}$,
from the same analysis.
\end{itemize}

It is important to underline that most of those estimates are {\it lower limits}.
As the
constraints derived on the Hubble parameter in the next sections would
have only been weakened by a lower CIB density, this makes the
following analysis conservative.
The {\it Spitzer} observatory data (Werner {\it et al.} \cite{werner}) allowed for deep
and wide area surveys,
in particular at 24, 70 and 160 $\mu$m using the MIPS
(Rieke {\it et al.} \cite{rieke}) which is a keypoint for this analysis. Those points, used here with the refined analysis of Dole {\it et al.} \cite{dole06},
improving over Dole {\it et al.} \cite{dole04}, are very relevant as they (at least two of them) lie within the
energy range where the interactions with gamma-rays are large. In the
following, we therefore consider the CIB density as known (within the
uncertainties) and use it as an "absorbing material" for the gamma-rays. The
higher the distance to the source ({\it i.e.} the lower the Hubble parameter), 
the higher the absorption. As the transmitted flux depends exponentially on the
depth of the absorber, the approach is very sensitive.\\

\begin{figure}
\includegraphics[width=90mm]{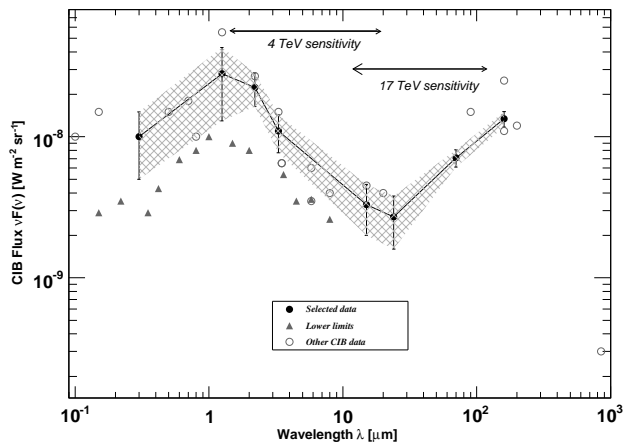}
 \caption{Cosmic infrared background energy density as a function of the wavelength.
 The filled circles are the data selected for this analysis. Open circles are tentative
 detections ignored in this article and triangles are lower limits. The intervals labeled 
 "4 TeV sensitivity" and "17 TeV sensitivity" correspond to 90\% of the interactions of,
 respectively, 4 TeV
 and 17 TeV gamma-rays taking into account the cross section convolved with the CIB
 spectrum. The shaded region corresponds to the $1\sigma$ allowed values of the CIB density, as included
  in the Monte-Carlo simulation.}
\label{fig1}
\end{figure}

Fig.~\ref{fig1} displays the data points (filled circles) used for this analysis together with 
lower limits (triangles) and estimates not taken into account as being affected by higher 
uncertainties (open circles). The Monte-Carlo method described in the next sections samples 
CIB realizations within the shaded region.

\section{Active galactic nucleus Mrk 501}

It has been suggested long ago that observations of the TeV spectrum of extragalactic sources can be a powerful
tool to constrain the CIB spectrum, especially around 10~$\mu$m, where the
Extragalactic Background (EB) 
constraints were quite weak. As a matter of fact,
TeV gamma rays propagating in the intergalactic medium 
undergo absorption through electron-positron pair production on CIB photons.
Using sources Mrk~421 and Mrk~501, meaningful upper limits have been 
established in different papers. More recently, interesting constraints were obtained by Aharonian {\it et al.}
\cite{aharon06}. They are based  on the observations performed with the very effective new network of
telescopes H.E.S.S. (Aharonian {\it et al.} \cite{aharon05}). In this 
article, we reverse the argument and use reliable estimates of the CIB density 
to obtain a new constraint on the Hubble parameter. The idea has already been pointed
out by Salamon {\it et~al.} \cite{salamon}. However, Mrk~501 data were not
available at that time and no result could be derived. Furthermore, the
method suggested by Salamon {\it et~al.} is far less conservative as it assumes that any rolloff in the
AGN spectrum should be due to the CIB absorption. This is a very strong
hypothesis as many different effects could (and, even, should) lead to a rolloff
intrinsic to the source. This could, among other processes, be due to 
self-absorption within the AGN, to the Klein-Nishina effect or simply to the
limiting energy available for Fermi acceleration to take place in a finite volume over a
finite amount of time. The bound derived in this article does not rely on this
"CIB rolloff" assumption and is therefore much more robust. But it can only lead to a
lower limit and not to a measurement.\\

The TeV source  Mrk~501 is the  second  closest X-ray selected BL~Lac 
object  after Mrk~421, with a redshift \mbox{$z = 0.034$.}
During the 1997 outburst which lasted  several months,   Mrk~501  was observed intensively
in X-rays (BeppoSAX:  Pian ~{\it et~al.} \cite{pian},  RXTE: Lamer \& Wagner \cite{lamer})    and  TeV 
$\gamma$-rays
(Whipple: Catanese ~{\it et~al.} \cite{catanese},   Samuelson ~{\it et~al.} \cite{samuelson};  
HEGRA: Aharonian ~{\it et~al.} \cite{aharon97},
Telescope Array: Hayashida ~{\it et~al.} \cite{hayashida};  
CAT:  Djannati-Ata\"{\i} ~{\it et~al.}  \cite{djannati}).  
The exceptional 1997 April~16 flare  was observed 
by  BeppoSAX and low-energy threshold ($\sim$~300~GeV)  
Whipple and   CAT atmospheric   Cherenkov telescopes,
allowing the derivation of the energy spectrum  with a good accuracy in a broad dynamical
range.
These unique data  initiated  interesting efforts to set meaningful   upper limits on the 
CIB flux (see {\it e.g.}  Biller ~{\it et~al.} \cite{biller},  Stanev \& Franceschini \cite{stanev}, 
 Stecker \& De Jager \cite{stecker98}, 
Aharonian {\it et~al.} \cite{aharon99b},  Coppi \& Aharonian \cite{coppi}, Konopelko ~{\it et~al.}
\cite{konopelko}, Guy ~{\it et~al.} \cite{guy},  Renault {\it et al.} \cite{renault}). 
More recently, other AGNs have been detected, especially with the H.E.S.S. network (see, {\it e.g.},
Aharonian {\it et. al.} \cite{aharon05b} for a review). As far as our study is
concerned, a promising opportunity is
the measurement of the spectrum of the distant blazar  1ES1101-232 (Aharonian {\it et~al.}
\cite{aharon07}). With a redshift $z=0.186$, this is a very good candidate to maximize the interaction
with the CIB. The spectrum has been measured by H.E.S.S. up to $\sim 2-3$~TeV. In fact, the optical
depth for gamma-rays is roughly the same at this energy for $z=0.186$ as at 
$\sim 15-20$~TeV for the lower Mrk~501
redshift $z=0.034$. The analysis hereafter described has also been performed for both
Mrk~501 and  1ES1101-232.
However, as the results are less stringent for the latter, we will mainly focus on Mrk~501.
Furthermore, the very wide range of measured energies (between $\sim$~400~GeV and $\sim$~20~TeV) for Mrk~501
allows for a much more reliable analysis. In this article, CAT and HEGRA data are used from 400~GeV to 21~TeV
(Aharonian ~{\it et~al.} \cite{aharon99b} and Guy {\it et al.} \cite{guy}). The
higher energy points, {\it i.e.} the more relevant ones for this analysis, come
from a time averaging of the HEGRA spectra based on more than 38000 detected
$\gamma$-rays. This very high number of data points combined with the 20\%
energy resolution of the instrument resulted in a very accurate measurement of
the spectrum above 10 TeV which remains unequaled.

\section{Interactions of gamma-rays with the infrared background}

The influence of low energy photons in the Universe on the propagation of 
Very High Energy (VHE) gamma-rays was first pointed out by Nikishov \cite{nikishov}. An original way of using
ground-based TeV observations of distant sources to probe the CIB was given by Stecker {\it et al.} 
\cite{stecker92}. The fundamental idea is to look for
absorption in the intrinsic spectrum as a result of electron-positron pair production by photon
collisions $\gamma_{TeV}+\gamma_{CIB} \rightarrow e^+ + e^-$.
In such an interaction between a gamma-ray of energy
$(1+z)E$ and an infrared photon of energy $(1+z)\epsilon$,  with
$E$ and $\epsilon$ the observed energies at $z=0$, the pair production threshold is
$E\epsilon\left(1+z\right)^2\left(1-cos\theta\right)>2\left(mc^2\right)^2$ where
$\theta$ is the angle between photons and $m$ the rest mass of the electron.
The cross-section can be written as (Heitler \cite{heitler}):
\mbox{$\sigma=\zeta\left(1-\beta^2\right)\left(2\beta\left(\beta^2
-2\right)+\left(3-\beta^4\right)ln\left(\frac{1+\beta}{1-\beta}\right)\right) {\rm
cm}^2$}
with 
$\beta=\left(1-2\left(mc^2\right)^2/(E\epsilon\left(1-cos\theta\right))\left(1+z\right)^2\right)^{1/2}$
and $\zeta=1.25\times10^{-25}$.
If the infrared photons have a density number $n\left(\epsilon\right)d\epsilon~{\rm
cm}^{-3}$, the corresponding optical depth for attenuation is
$$
\tau(E) \approx \frac{c z_s}{H_0}
\int_{-1}^{1}d\left(cos\theta\right)\frac{1-cos\theta}{2}
\int_{\epsilon_t}^{\infty}d\epsilon \ n(\epsilon)\sigma(E,\epsilon,\theta)
$$

 where $\epsilon_t
=2(mc^2)^2/(E(1-cos\theta)(1+z)^2)$, $z_s$ is the redshift of the source and $c$ the
speed of light. This expression is strictly valid for an Einstein-de Sitter universe but remains an
accurate approximation in the $\Lambda CDM$ paradigm for low redshifts. The detected flux is then attenuated by a
factor of $e^{-\tau(E)}$. The CIB energy distribution is assumed to be independent of $z$ as the $\gamma$-ray source redshift 
is very low (0.034).
The maximum cross-section is reached for an infrared photon
wavelength of $\lambda_{CIB}\approx \lambda_c \frac{E}{2mc^2}$ where
$\lambda_c=h/(mc)$ is the Compton wavelength of the electron. Therefore,
$\gamma$-photons with energy between a few TeV and
20~TeV "see" CIB photons with wavelengths between 3.5 and 100~$\mu$m.\\

To allow for an intuitive understanding of the exponential absorption (not only as a function of the
quantity of the absorbing material, as it should be, but also as a function of the energy of the gamma-rays), one can assume 
that the
CIB is roughly constant, {\it i.e.} $\epsilon^2 n(\epsilon)\approx cte$. The optical
depth for an energy $kE$ can then be written as:
$$\tau(kE)\propto \int_{-1}^{1}d\left(cos\theta\right)\frac{1-cos\theta}{2}
\int_{\frac{2\left(mc^2\right)^2}{kE\left(1+z\right)^2\left(1-cos\theta\right)}}^{\infty}d\epsilon
\frac{\sigma(kE\epsilon,\theta)}{\epsilon^2}.$$
By a change of variable $\epsilon \rightarrow k\epsilon$, it is straightforward to see that
$\tau(kE) = k\times\tau(E)$. As $\Phi=\Phi_0\times e^{-\tau(E)}$, this explains why a high energy
gamma-ray will be exponantially more absorbed than a low energy one.

\section{Unfolding method and physical criteria}

To turn the absorption corrected spectra into a lower limit on the Hubble parameter, the intrinsic
spectral energy density of Mrk~501 is assumed to be concave ({\it i.e.} with a negative
second derivative) in the multi-TeV region.
This quite common approach has been developed in Renault {\it et al.} \cite{renault}
and Guy {\it et al.} \cite{guy}. 
The conservative hypothesis is based on the fact that no natural physical process can
re-inject energy above the Inverse-Compton bump maximum. Both Klein-Nishina effect and 
auto-absorption within the source would only cause the flux to decrease more and more rapidly
as a function of energy. This is obviously true for leptonic models (a good review of theoretical
perspectives, where each model produces a concave spectrum, can be found in Saug\'e \cite{sauge}) but also for
more complex hadronic models where this
concave shape is also expected, either because of proton-initiated-cascade (often modeled as a broken
power-law around 3 TeV), or due to the inclusion of $\mu$-synchrotron radiation (Mannheim, 
private communication). Even in the extreme case where proton synchrotron radiation is at the origin of 
the TeV bump (Aharonian \cite{aharon00}), the shape of the emission remains concave.
Looking at the absorption corrected spectra, it cannot be totally excluded that the maximum of 
the so-called Inverse-Compton peak is not yet reached at 20~TeV. Physical parameters (essentially
magnetic field and Doppler factor) required to produce such a spectral energy distribution maximum 
above 17 TeV are substantially disfavored but, even in this case, the $\nu F(\nu)$ shape should 
remain concave. This latter point is demonstrated by BeppoSAX 1-100~keV
measurements (Pian {\it et al.} \cite{pian}) showing a
clearly concave spectrum before the synchrotron bump maximum 
(around 100~keV) which is supposed to be mimicked by the
TeV-spectrum before the so-called Inverse-Compton bump maximum. This hypothesis is reinforced by the fact
that the sub-TeV slope, which is independent of the CIB density beyond 3.5~$\mu$m,
effectively reflects the X-ray slope in the keV range.
This behavior (together with the correlated variability reported {\it e.g.} by Aharonian {\it et al.} 
\cite{aharon99a}
and Djannati-Ata\"{\i} {\it et al.} \cite{djannati}) indicates that the same population of particles  is at the origin 
of both X-ray and $\gamma$-ray emissions, whatever this population is. In particular, the 
self-synchro-Compton model fits satisfactorily 
the absorption corrected data (Guy {\it et al.} \cite{guy}).\\

Furthermore, in addition to those "theoretically motivated" arguments, observations
show that the spectrum should indeed be concave. This can be checked by considering AGNs
with an IC bump seen at lower energies and therefore unaffected by the CIB absorption.
This can easily be seen for 3C273 with the average spectrum compiled from 30 years of
observations as reported in T\"urler {\it et al.} \cite{turler}. This is also true
for Cen A (see, {\it e.g.}, Chiaberge {\it et al.} \cite{chiaberge}) which is the 
nearest radiogalaxy (z=0.0018) and one of the best studied. Although with less
significance this can even be checked for M 87 (which is not a blazar), as shown in
Lenain {\it et al.} \cite{lenain}. The FSRQ PKS 0521-36 (see, {\it e.g.} the summary
presented in Lenain {\it et al.} \cite{lenain}, mostly from Giommi {\it et al.}
\cite{giommi} and the NED), which is known for oscillating
between a Seyfert-like and a BL Lac state, also exhibits a fully concave spectrum. This
is also true for the blazar 3C 454.3, as measured during the 2007 July flare (Ghisellini
{\it et al.} \cite{ghi}). During flares, good quality data reinforcing the concavity
assumption are also available for PKS
2155-304 (Aharonian {\it et al.} \cite{aharo07}) and for 3C 279 (see, {\it e.g.}, the
summary plot in B\"ottcher {\it et al.} \cite{bott}).\\

It should be stressed that requiring the spectrum to be concave is "more demanding" than
simply requiring the end of the TeV spectrum not to exhibit a too hard spectral index,
as often used to derive limits on the CIB. Our approach evades the  argument of Stecker 
{\it at al.} \cite{stecker07} questioning the results of Aharonian {\it et al.} \cite{aharon06}.\\

\begin{figure}
\includegraphics[width=84mm]{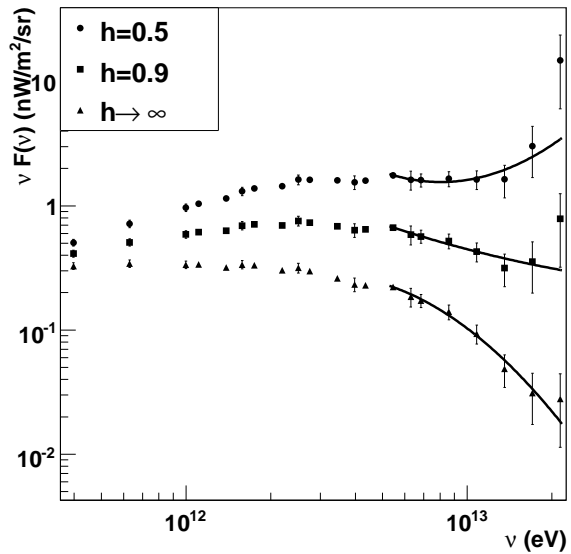}
 \caption{Unfolded gamma-ray spectra and associated fits for (from bottom to top) : no CIB (equivalent to
 $H_0\rightarrow \infty$),  $H_0=90~{\rm km}.{\rm s^{-1}}.{\rm Mpc}^{-1}$ and $H_0=50~{\rm km}.{\rm s^{-1}}.{\rm Mpc}^{-1}$.}
\label{fig2}
\end{figure}

In order to quantify the concavity of the absorption corrected spectrum, 
a parabolic fit is performed from 6 to 21~TeV 
in the plane ($log(\nu),log(\nu F \nu))$. 
This function, which is simple, is chosen for its constant second derivative 
$a= d^2(log(\nu F \nu))/d^2(log(\nu))$ which
avoids the choice of a particular test-energy and fits satisfactorily the data. Over such a
small interval it is obviously meaningful to assume the spectrum to be locally parabolic.
The previous physical constraint on the concavity of the TeV spectrum simply
reads as $a<0$. As explained in the next section, the uncertainties are carefully taken
into account.
The parameter $a$ is computed for different values of the Hubble parameter and
the corresponding value is rejected if $a$ is positive within the associated
errors. 
Fig.~\ref{fig2} shows the result of the parabolic fit in several cases, superimposed with the
experimental
points. The fit under-estimates the energy density of the hardest photons,
ensuring  that the local second derivative is under-estimated, thus making the test conservative.\\

\section{Results on the Hubble parameter}

To derive statistically meaningful constraints on the Hubble parameter, errors on both the CIB and the 
measured TeV spectrum have been taken into account. As far as the gamma-ray spectrum is concerned, we
have not only considered the statistical uncertainties but also the systematic errors, as evaluated by
the HEGRA team (Aharonian {\it et al.} \cite{aharon99b}). The errors on the CIB density are not as easy
to take into account as those on the AGN spectrum as their propagation is far from trivial. For this
reason, we have performed a full Monte-Carlo simulation. Each of the eight considered CIB points between
0.3 and 160 $\mu$m is randomly generated according to its central value and
uncertainty, as displayed by the shaded region of Fig.\ref{fig1}. Linear interpolations are 
then performed in the log-log plane to obtain a CIB realization. For each CIB
realization, the unfolded spectrum is computed ({\it i.e.} the optical depth is numerically evaluated for the
considered energies with the formula given in section 3) and fitted with a parabola between 6 and 21 TeV. 
The error $\Delta a$ on the second derivative $a$ of the parabola (due to uncertainties in the
measurement of the gamma-ray spectrum) is computed according to three different
statistical methods iterated up to the convergence point. The value of the
Hubble parameter is rejected at $x\sigma$ if $a-x\Delta a>0$. The procedure is finally repeated to
generate many (typically $\sim 1000$) realizations of the CIB. Confidence levels can be computed in this way 
by estimating the percentage of rejected realizations for each tested 
value of the Hubble parameter. It has been checked that our results are marginally dependent upon the
number of points used in the fit and that the $\chi^2$ per degree of freedom is, in each case, low
enough to ensure a satisfactory fit quality.\\

At the 68\% confidence level, the lower limit derived on $H_0$, taking into account all the
uncertainties is: $H_0>74~{\rm km}.{\rm s^{-1}}.{\rm Mpc}^{-1}$. This is the main result of this
article. Although it should be underlined that even if a 1$\sigma$ limit remains
rather weak on a statistical basis, this bound is meaningful --at this confidence
level-- when compared with other results. The history of the measured value of
the Hubble parameter is very chaotic and different (incompatible) values have
been taken as granted at different times. As reminded in the recent review 
article "the Hubble constant" by Jackson \cite{jack}, the current estimates range between
60 and 75 ${\rm km}.{\rm s^{-1}}.{\rm Mpc}^{-1}$. In this context, our lower
limit $H_0>74~{\rm km}.{\rm s^{-1}}.{\rm Mpc}^{-1}$ is unquestionably relevant.\\

As far as Cepheids are concerned, the main measurements are $H_0=73~{\rm km}.{\rm s^{-1}}.{\rm Mpc}^{-1}$
(Riess {\it et al.} \cite{riess}), with statistical errors of $4~{\rm km}.{\rm s^{-1}}.{\rm
Mpc}^{-1}$ and systematic errors of $5~{\rm km}~{\rm s^{-1}}~{\rm
Mpc}^{-1}$ on the one hand and $H_0=62.3~{\rm km}.{\rm s^{-1}}.{\rm Mpc}^{-1}$
(Sandage {\it et al.} \cite{sandage}), with statistical errors of $1.3~{\rm km}~{\rm s^{-1}}~{\rm
Mpc}^{-1}$ and systematic errors of $5~{\rm km}.{\rm s^{-1}}.{\rm
Mpc}^{-1}$ on the other hand. Our results clearly disfavor the Sandage estimate, even within its
quoted uncertainty. As far as gravitational lenses are concerned, a sophisticated
meta-analysis has recently been performed (Oguri {\it et al.} \cite{oguri}) using Monte-Carlo methods
to account for quantities such as the presence of clusters around the main lens
galaxy and the variation in profile slopes. The result obtained is 
$H_0=68\pm6\pm8~{\rm km}.{\rm s^{-1}}.{\rm Mpc}^{-1}$. Meanwhile, an approach
modelling simultaneously 10 of the 18 time-delay lenses with non-parametric
models, has led (Saha {\it et al.} \cite{saha}) to $H_0=72\pm12~{\rm km}.{\rm s^{-1}}.{\rm
Mpc}^{-1}$. Our results allow a  significant $1\sigma$ narrowing of those
uncertainty intervals. As far as the Sunyaev-Zeldovich effect is concerned, recent
measurements (see, {\it e.g.} Jones {\it et al.} \cite{jones}) are compatible with those estimates
but the errors are too large to allow for any improvement.\\

The most reliable measures available to date are probably
those reported in the 
"final results from the Hubble Space Telescope Key Project" (Freedman {\it et al.} \cite{free}). They
concluded that the full analysis of all data gives $H_0=72\pm 8~{\rm km}.{\rm s^{-1}}.{\rm
Mpc}^{-1}$. Combining this HST
allowed interval with our lower limit increases the mean value to $H_0\approx 76~{\rm
km}.{\rm s^{-1}}.{\rm Mpc}^{-1}$. However, the statistical significance of this estimate should
be taken with care as the accurate distribution is not known. Nevertheless, our result 
$H_0>74~{\rm km}.{\rm s^{-1}}.{\rm Mpc}^{-1}$ clearly favors the upper
end of the HST interval and substantially reduces the allowed parameter space.
A less constraining, although
interesting, result was obtained by Chandra (Bonamente  {\it et al.}
\cite{bona}) leading to $H_0=77\pm 12~{\rm km}~{\rm s^{-1}}~{\rm
Mpc}^{-1}$ with a larger 15~\% uncertainty. Finally, WMAP has published in the five-years results
(Komatsu {\it et al.} \cite{komatsu}) an impressive estimate at $H_0=72\pm 2.6~{\rm km}~{\rm s^{-1}}~{\rm
Mpc}^{-1}$. It should, however, be strongly emphasized that
WMAP results rely on many different priors and do not allow for an independent
determination of $H_0$ due to its degeneracy with the total curvature of the Universe. For
example, every decrease of $20~{\rm km}.{\rm s^{-1}}.{\rm Mpc}^{-1}$ in $H_0$ increases the
total density of the Universe by 0.1 in units of the closure density. The WMAP data by
themselves, without any further assumptions or extra hypothesis, do not supply a significant
constraint on $H_0$.\\

If the statistical significance is to be
improved to 90\%, the lower limit derived in this work is significantly reduced to $H_0>56~{\rm
km}.{\rm s^{-1}}.{\rm
Mpc}^{-1}$. This is, of course, due to large uncertainties on the CIB density.
This broadening is somehow analogous to what happens for other approaches to
measure $H_0$. For example, Tegmark {\it et al.} \cite{teg} have shown that when
relaxing the constraint of the equation of state of the dark energy, the 90\%
confidence level estimate using WMAP data is $H_0 \in [61-84]~{\rm km}.{\rm s^{-1}}.{\rm Mpc}^{-1}$.

\section*{Conclusion and prospects}

\begin{figure}
\includegraphics[width=84mm]{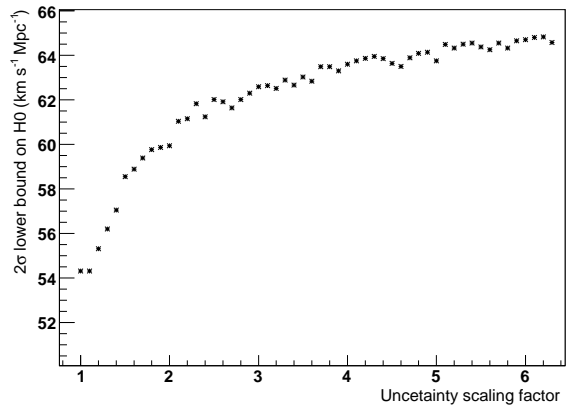}
 \caption{Lower limit at the 90\% confidence level on the Hubble constant as a function of the rescaling factor of
 the CIB density. A factor of $x$ means that the errors are assumed $x$ times smaller than
 currently estimated. Each point is obtained by analysing 1000 Monte-Carlo realizations of the
 CIB.}
\label{fig3}
\end{figure}

The lower limit on $H_0$ derived in this letter is totally {\it independent} of the other available
measurements. It is conservative, both due to the method which tends to 
under-estimate the high energy flux density and to the weak hypothesis regarding the spectrum concavity. 
Internal effects, such as Klein-Nishina cutoff and self-absorption, lead to very sharp multi-TeV intrinsic
spectra. Furthermore, and most importantly, the CIB density is underestimated as only lower limits are 
used. Nevertheless, the errors on those lower limits have been exhaustively taken into account thanks to
a Monte-Carlo simulation.
Our result allows a substantial $1\sigma$ narrowing of previously obtained uncertainty intervals, favoring the
higher possible values of $H_0$. It also closes the window on exotic "low Hubble constant" scenarios
(Nugent {\it at al.} \cite{nugent}) and on some alternatives to the cosmological concordance model (see,
{\it e.g.}, Blanchard {\it et al.} \cite{blanchard} where $h\approx 0.46$ is assumed.)\\

Clearly, this work aims at providing a
first hint in this direction and a first order of magnitude of what can be
obtained by this method. Although the statistical significance of our results has been computed with
care, several improvements could be expected in the future. First, other AGNs will probably be detected
by H.E.S.S (Aharonian {\it et al.} \cite{aharon05}), MAGIC
(Bastieri \cite{bigo}) or VERITAS (Krennrich \cite{kre}) and should be included
in the analysis. GLAST (Latronico {\it et al.} \cite{latro}) data, at lower energies, could even be useful. This is very promising to account for more sources at different redshifts. However, the
current situation for known AGNs is not expected to be improved significantly for a simple reason: the most
important part of the spectrum used to constrain the Hubble parameter is the high-energy tail and larger
telescopes will not improve the sensitivity in this range (just because the effective area is determined
by the area of the shower and not by the area of the mirror). Then, improvements can be
expected from measurements of the CIB. Fig~\ref{fig3} displays the evolution of the
$2\sigma$ lower limit as a function of the CIB uncertainty scaling factor. A scaling factor
$x$ means that the errors are assumed to be $x$ times smaller than currently estimated. The lower
bound is, as expected, improved by smaller uncertainties and reaches a limit when the
propagation of errors induced by the CIB uncertainties becomes much smaller than the effect
of gamma-ray uncertainties associated with the measurement of the AGN spectrum. For each
point of the plot, 1000 Monte-Carlo realizations have been computed and analyzed according to
the method described in the previous section.\\

In the near future, measurements should indeed be improved between 60 and 110
$\mu$m thanks to {\it Herschel} (see, {\it e.g.} Franceschini {\it et al.} \cite{frances}) and, in the far future, between 5 
and 60 $\mu$m thanks to {\it JWST} (see, {\it e.g.} Windhorst {\it et al.} \cite{wind}). This could make this
approach quite competitive.\\

Finally, if gamma telescopes become so efficient that many more blazars are detected, 
each of them exhibiting a redshift-dependant cutoff in the spectrum, then the approach 
suggested by Salamon {\it et al.} \cite{salamon} could be tractable and would lead not only to a 
bound but to a measurement of the Hubble parameter. However, to distinguish between a CIB 
absorption effect and an intrinsic cutoff, a very large number of data points would be required.

\label{lastpage}

\end{document}